\documentclass[%
 reprint,
 amsmath,amssymb,
 aps,
]{revtex4-2}

\usepackage{xcolor}
\usepackage[normalem]{ulem}
\usepackage{graphicx}

\usepackage{hyperref}
\hypersetup{
    colorlinks=true,
    citecolor=black,
    linkcolor=black,
    urlcolor=blue,
}

\begin{document}

\title{Physical limitations on broadband invisibility based on fast-light media}

\author{Mohamed Ismail Abdelrahman}
\thanks{These authors contributed equally}
\affiliation{%
	Cornell University, School of Electrical and Computer Engineering, Ithaca, New York 14853, USA
}%

\author{Zeki Hayran}
\thanks{These authors contributed equally}
\affiliation{%
	Cornell University, School of Electrical and Computer Engineering, Ithaca, New York 14853, USA
}%

\author{Aobo Chen}
\affiliation{%
	Cornell University, School of Electrical and Computer Engineering, Ithaca, New York 14853, USA
}%

\author{Francesco Monticone}
\thanks{francesco.monticone@cornell.edu}
\affiliation{%
	Cornell University, School of Electrical and Computer Engineering, Ithaca, New York 14853, USA
}%

\maketitle

This note is a comment on a recent article  \cite{2019Tsakmakidis} that presents a thought-provoking proposal to overcome the bandwidth restrictions of invisibility cloaks based on using media that support superluminal (faster than light in free space) group and phase velocities. 
As illustrated in Fig. 1 of Ref. \cite{2019Tsakmakidis}, a wave packet propagating through such a  “fast-light cloak” is alleged to be able to reach the side behind the cloaked object simultaneously with a corresponding wave packet propagating through the shorter, direct route in free space without the object, so that “no shadow or waveform distortion arises.” As the authors claim, the “extra pathlength is balanced out by the correspondingly larger group velocity of the pulse in the cloak”, which allows to “restore the incident field distribution all around the object in, both, amplitude and phase”. This fast-light effect may be achieved in a broadband fashion using active (gain) materials. The authors claim that such a “fast-light cloak” can hide an object, even from time-of-flight detection techniques, and achieve invisibility “over any desired frequency band, so long as the superluminality condition [...] is attained over the desired bandwidth.” 
We disagree with these claims and believe that a thorough clarification of the ideas put forward in Ref. \cite{2019Tsakmakidis} is important and necessary for the broad wave-physics community. Specifically, in this comment we clarify that invisibility cloaks based on fast-light media suffer from fundamental bandwidth restrictions that arise due to causality, the nature of superluminal wave propagation, and the stability issues of active systems. These limitations and issues were not addressed in Ref. \cite{2019Tsakmakidis}. Most importantly, we show that the material model considered in Ref. \cite{2019Tsakmakidis} is unphysical.

It is a well established fact that relativistic causality (signals cannot travel faster than light in vacuum) allows perfect invisibility only over a zero-measure bandwidth, as conclusively demonstrated in Ref. \cite{2006Miller}. In addition, for a specified level of acceptable wavefront distortion and scattering, causality sets an upper bound on the bandwidth over which an imperfect invisibility effect can be obtained  \cite{craeye2012rule,2010Hashemi,2016Monticone}. The authors of Ref. \cite{2019Tsakmakidis} claim to achieve ``true invisibility,'' ``over any desired frequency band, so long as the superluminality condition [...] is attained over the desired bandwidth,'' realizing cloaking devices that ``cannot be detected using interferometric or time-of-flight techniques.'' 
%
While the authors do not explicitly claim to achieve ``perfect'' invisibility (identically zero scattering cross section), we feel that the claims of Ref. \cite{2019Tsakmakidis} quoted above -- especially in relation to bandwidth --  may introduce significant confusion regarding the role of causality and fast-light media in the context of invisibility and cloaking. For these reasons, in the first part of this comment, we would like to take the opportunity to clarify whether fast-light media (and, more generally, active media) can circumvent causality limitations, providing a concise review of this issue. 


A crucial fact of wave physics is that the group velocity of a wave packet in a physical medium is not, in general, equivalent to the energy and information velocity. While the latter is bounded by the speed of light in vacuum \cite{1960Brillouin} (relativistic causality), nothing prohibits the peak of a smooth pulse from propagating superluminally in fast-light media, corresponding to superluminal group velocity (SGV), as shown in Fig. 1 for the case of one-dimensional SGV propagation. However, it must be understood that the superluminal peak emerging from a fast-light medium is merely a result of reshaping of the input pulse \cite{1970Garrett,1993Chiao,1996Diener,1998Garrison}. Although the authors of Ref. \cite{2019Tsakmakidis} have correctly acknowledged the distinction between information velocity and group velocity, they failed to recognize that the information velocity fundamentally matters in the detection process and sets the ultimate upper bound for the cloaking bandwidth for a specified acceptable level of distortion and scattering \cite{craeye2012rule,2010Hashemi,2016Monticone}.
%
The matching of the peak of the pulse, which might propagate superluminally, with its free-space-propagating counterpart is generally insufficient to obtain a cloaking effect that is fully robust to time-of-flight or interferometric measurements. 


Information velocity has been associated, as early as in 1907 by Sommerfeld 
\cite{1960Brillouin}, with the  “front discontinuity” of a pulse (the discontinuity due to the switching-on of the generator/transmitter), which contains the highest spectral components of the signal \cite{1997Chiao}. In a more
general sense, the appearance of  ‘genuine’ information can only be represented by non-analytic points in the function representing the pulse
shape (envelope)  \cite{1996Diener,1997Chiao,1998Garrison}. Irrespective of the background medium, the front discontinuity propagates at the speed of light in vacuum, the upper bound set by Einstein’s relativistic causality principle \cite{1960Brillouin,1997Chiao}, as illustrated by the example in Fig. 1. One may argue that the aforementioned definition of information velocity based on a discontinuity is merely a  mathematical idealization, and that the pulse front is typically too small to be detected. 
In this context, Kuzmich et. al. \cite{2001Kuzmich} and Stenner et. al. \cite{2003Stenner} have proposed a “practical” definition for information velocity in terms of the bit-error-rate required by the detection system to “announce” the arrival of new information, that is, the detection of a new symbol, for example “0” or “1”. The detection time is defined as the time instant at which the bit-error-rate of the received signal drops below a certain, arbitrary, threshold. 
Recently, this  definition has been termed the “velocity of detectable information” \cite{2014Dorrah}. Within this context, it has been theoretically and experimentally verified that information carried by a pulse propagating in a fast-light medium is always received by a sufficiently efficient detector after the information carried by a companion pulse propagating in free space for the same distance \cite{2001Kuzmich, 2003Stenner,2014Dorrah}. These results further confirm that superluminal group velocity is not equivalent to the strictly forbidden superluminal information velocity. The retardation compared to free-space is unavoidable as it ultimately originates from the pulse distortion and the intrinsic quantum noise of the gain process in fast-light media \cite{2001Kuzmich,2010Boyd}. 
In particular, it is crucial to note that pulse distortion in fast-light media is inevitable, as there are always some spectral components of the input signal that lie outside the band over which the medium supports superluminal group velocity (the spectrum of a real time-limited signal with an arbitrary switching-on front is never finite, i.e., a time-limited signal is not band-limited) \cite{2013Hrabar}. Consequently, fast-light media or, more generally, any media, cannot advance the detection process compared to the source-detector system in free space. This is in contradiction with the suggestion made in Ref. \cite{2019Tsakmakidis} that fast-light media ``lead to breaking scattering causality, where the presence of a scatterer (object) causes the wave to reach the detector faster than in the scatterer’s absence.'' Instead, as discussed above and repeatedly shown in the literature, a sufficiently good detector would always be able to distinguish between a non-monochromatic electromagnetic pulse propagating through a fast-light medium and one propagating in free space. 
%
%
As a result, a cloaked object of any size would, in principle, always be detectable using time-of-flight techniques, even if fast-light media are employed. We reiterate that, unlike the practical limitations that might arise due to imperfect implementations, relativistic causality is a fundamental constraint preventing perfect invisibility over any non-zero bandwidth \cite{2006Miller}. For imperfect invisibility, causality determines an upper bound on the bandwidth of the cloaking effect, for a certain level of acceptable wavefront distortion \cite{craeye2012rule,2010Hashemi} (which translates into a certain scattering cross section), with the bandwidth bound becoming more stringent if the cloaking device is assumed to be passive \cite{2016Monticone}.

While fast-light media cannot achieve superluminal information transfer, they still have intriguing applications. For example, they can be used to reduce the detection latency, mainly when the noise of the detector is predominant compared to the noise of the gain medium \cite{2014Dorrah}. In the context of cloaking, gain media, fast-light media, and so-called non-Foster circuits (active circuits implementing negative resistances, capacitances, inductances) have already been proposed, years before Ref. \cite{2019Tsakmakidis}, to realize invisibility cloaks that, while not perfect and still constrained by causality, are indeed more broadband and efficient than their passive counterparts \cite{2013Chen,2019Aobo}. 
Yet, there are several limits and design considerations that must be taken into account and deserve further discussion. For instance, there are bounds on the size of a fast-light medium that define the approximately-undistorted propagation regime in which SGV could be useful. These bounds depend on the material parameters and the pulse bandwidth \cite{1993Chiao,2014Dorrah}. In addition, a superluminal pulse peak can never overtake the pulse front  \cite{1998Garrison}.

Most importantly, whenever active media are used to design invisibility cloaks, or any other device, it is always critically important to carefully assess their stability, namely, the absence of unbounded oscillations in their temporal impulse response (unstable poles in their transfer function). Indeed, the complex wave interactions inside a cloaking structure containing active media may trigger unstable field oscillations under a generic illumination (or just noise), transforming the invisibility cloak into an emitting (lasing) structure, thereby disrupting the entire cloaking functionality. Stability can be ensured with a proper design, but usually at the cost of limiting the operational bandwidth of the cloak and the maximum scattering suppression, as we discuss in \cite{2019Aobo}. 
However, no discussion of stability, or of the transient temporal response of the invisibility cloak, is provided in Ref. \cite{2019Tsakmakidis}. 

In fact, a simple inspection of the material dispersion model of the proposed cloak in Fig. 2 of Ref. \cite{2019Tsakmakidis} (see Eqs. (1),(2) in the Methods section) reveals that the hypothetical electric and magnetic susceptibilities of the cloaking material are actually non-causal (despite the authors' claim of having performed “full-wave causal simulations”) and, therefore, the proposed cloak is not physically realizable. 
Indeed, the material models considered in Ref. \cite{2019Tsakmakidis} exhibit a non-zero (negative) high-frequency limit for the susceptibilities (as shown in Fig. 2a), and hence a less-than-unity limit for the refractive index $n$. This behavior clearly violates relativistic causality because it directly implies that the front of the pulse, which is composed of the highest frequency components of the signal, would propagate superluminally, corresponding to superluminal information velocity. Quoting Ref. \cite{2006Fearn}: ``Recent articles on superluminal signal propagation based on unphysical models which allow $n(\omega) \rightarrow \beta$ when $\omega \rightarrow \infty$ where $\beta < 1$ must therefore be dismissed as a violation of relativistically causal behavior.'' A more intuitive way to see the unphysical nature of this model is the following. A system (e.g., a bound electron) subject to a finite oscillatory driving force with a frequency much higher than all of the system's resonant frequencies will not be able to react to that force due to its inertia (this is also the case for a simple driven pendulum); therefore, any material susceptibility -- originating microscopically from an ensemble of harmonic oscillators -- needs to converge to zero for diverging frequency to ensure a physical response \cite{1960Brillouin}. Notably, if one were allowed to choose the high-frequency limit of the susceptibilities arbitrarily (and, thus, ignore any implications of relativistic causality), a similar low-loss, broadband, fast-light regime could be implemented even without resorting to any gain media, which further confirms the unphysical nature of this material model.

In principle, one could make the material dispersion considered in Ref. \cite{2019Tsakmakidis} causal, while maintaining the same optimized refractive index profile in the frequency range of interest, for example through the addition of a third inverted Lorentzian resonance with suitable parameters, ensuring that the susceptibility vanishes for diverging frequency (see Fig. 2a). 
In particular, the resonance frequency $\omega_0$ and the ratio of plasma frequency $\omega_p$ to $\omega_0$ for this additional resonance should be large with respect to the original two inverted Lorentzian resonances, in order for the refractive index of the cloaking material to remain approximately dispersionless and between zero and unity in the frequency range of interest. This modified, causal, material dispersion, however, typically leads to an unstable scattering response \cite{2019Aobo}, characterized by the onset of unbounded temporal oscillations (corresponding to unstable scattering poles), as shown in Fig. 2c,d and Supplementary Video 1$^{\dag}$. 
Although stability may be ensured by suitably tuning the parameters, for example by decreasing $\omega_0$ and $\omega_p/\omega_0$, the refractive index would then become substantially dispersive within the frequency band of interest, which would compromise the cloaking bandwidth considerably. In addition, stability also typically demands a broad linewidth for the inverted Lorentzian resonances \cite{2019Aobo}, which implies a larger imaginary part of the refractive index within the operating bandwidth, thereby further deteriorating the cloaking performance. As a result, the cloaking bandwidth and the maximum scattering suppression would be limited if causality and stability are both properly ensured, as was shown in Ref. \cite{2019Aobo} with physical material dispersion models. Consequently, one cannot simply conclude that an invisibility cloak based on fast-light media can operate “over any desired frequency band, so long as the superluminality condition is attained,” as claimed in Ref. \cite{2019Tsakmakidis}, without first performing a careful time-domain and stability analysis, and making sure that the proposed material model does not violate fundamental physical laws. 


In conclusion, we believe this comment fully clarifies the potential and limitations of invisibility cloaks made of fast-light media, addressing some of the issues related to the claims and results of Ref. \cite{2019Tsakmakidis}.
While active cloaks do have the potential to relax some of the limitations of passive devices, as already discussed in several earlier papers, the causality and stability of the cloaking system must always be carefully assessed and enforced. In summary, we believe that the idea of using fast-light (and, more generally, active) media to enhance the performance of invisibility cloaks is interesting and might have important implications in the future, but only if all physical limitations in such systems are properly considered and acknowledged.

\section*{Methods}
\textbf{Simulations.} All simulations were performed in time-domain via the finite-difference time-domain method using a commercially available software (Lumerical FDTD Solutions). In Fig. 1, the pulses have a center frequency of $562$ THz and a bandwidth of $210$ THz. The slab exhibits an average group velocity of $5.5 c_0$ over the bandwidth of interest ($c_0$ is the velocity of light in vacuum), and has a thickness of $22$ nm. 

In Fig. 2, the spherical object to be cloaked has a radius of 90 nm and a relative permittivity of 5, while the thickness of the spherical shell cloak is equal to 10 nm. The electric and magnetic susceptibilities of the cloak material follow standard multi-resonator dispersion models, as in Eqs. (1), (2), respectively:

%

\begin{equation}
\chi_\textrm{e}(\omega)=\varepsilon_\infty -1 + \sum_{n=1}^{m} \frac{f_n\omega^2_{\textrm{p1},n}}{\omega^2_{0,n} - \omega^2 + i\omega\gamma_n}
\end{equation}

\begin{equation}
\chi_\textrm{m}(\omega)=\mu_\infty -1 + \sum_{n=1}^{m} \frac{f_n\omega^2_{\textrm{p2},n}}{\omega^2_{0,n} - \omega^2 + i\omega\gamma_n}
\end{equation}
\vspace{3mm}

The material dispersion parameters of the original “fast-light cloak” (Fig. 2 of Ref. \cite{2019Tsakmakidis}, which purportedly included “time-domain simulations snapshots”) are listed below (as provided by the authors through private communications), which correspond to a non-causal gain doublet configuration with two inverted Lorentzian resonances (Fig. 2a):

$\varepsilon_\infty = 0.12$, $\mu_\infty = 0.06$, m = 2, $f_1 = f_2 = -1$, $\omega_{\textrm{p1},1} = \omega_{\textrm{p2},1} = 0.05\omega_\textrm{c}$, $\omega_{p1,2} = \omega_{p2,2} = \omega_\textrm{c}/12$, $\omega_{0,1} = 0.25\omega_\textrm{c}$, $\omega_{0,2} = 3\omega_\textrm{c}$, $\gamma_1 = \gamma_2 = 0.06\omega_\textrm{c}$, $\omega_\textrm{c} = 2\pi500$ THz 
\vspace{3mm}

The parameters of our causal dispersion models are listed below, where an additional inverted Lorentzian resonance has been added to both the electric and magnetic susceptibility of the original fast-light material to obtain a causal response (Fig. 2a).

$\varepsilon_\infty = 1$, $\mu_\infty = 1$, m = 3, $f_1 = f_2 = f_3 = -1$, $\omega_{\textrm{p}1,1} = \omega_{\textrm{p}2,1} = 0.05\omega_\textrm{c}$, $\omega_{\textrm{p}1,2} = \omega_{\textrm{p}2,2} = \omega_\textrm{c}/12$, $\omega_{\textrm{p}1,3} = 18.762\omega_\textrm{c}$, $\omega_{\textrm{p}2,3} = 19.391\omega_\textrm{c}$, $\omega_{0,1} = 0.25\omega_\textrm{c}$, $\omega_{0,2} = 3\omega_\textrm{c}$, $\omega_{0,3} = 20\omega_\textrm{c}$, $\gamma_1 = \gamma_2 = 0.06\omega_\textrm{c}$, $\gamma_3 = \omega_\textrm{c}$, $\omega_\textrm{c} = 2\pi500$ THz
\vspace{3mm}

The scattering spectra and scattering coefficient in Fig. 2b,c were obtained using exact Mie theory \cite{Bohren}. In Fig. 2d and the supplementary animation, the pulse has a center frequency of 650 THz and a bandwidth of 500 THz.


\begin{description}
\item[Competing Interests]
The authors declare no competing interests.

\item[Data availability]
All relevant data are available from the corresponding author upon reasonable request.

\item[Author contributions]
F.M. initiated and supervised the work. M.I.A. led the writing of the manuscript, with revisions from the other authors; Z.H. and A.C. performed the calculations and simulations; Z.H. prepared the figures and the supplementary videos; All the authors contributed to extensive discussion of the project.

 \item[Correspondence] 
 Correspondence and requests for materials
should be addressed to F.M. (email: francesco.monticone@cornell.edu)
\end{description}


\newpage




\begin{figure*}
\centering
  \includegraphics[ width=0.5\linewidth, keepaspectratio]{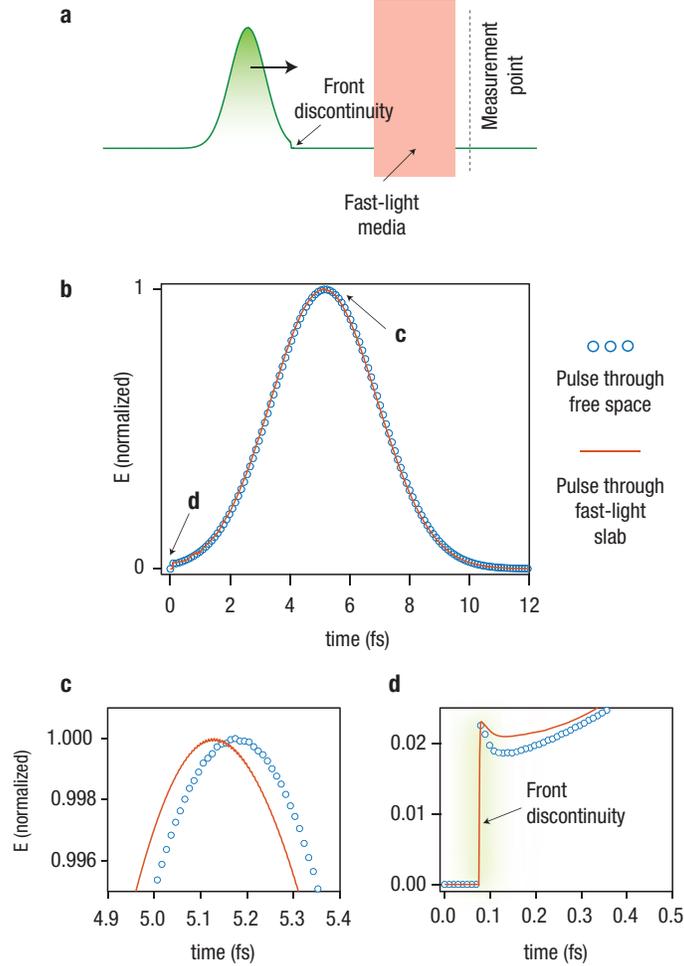}
\caption{\textbf{Physical limitations on superluminal pulse propagation due to relativistic causality}. (a) Conceptual scheme of a broadband pulse (envelope of a wavepacket) propagating through a slab composed of a fast-light medium. (b) Transmitted electric-field envelope, recorded in time using fully causal FDTD simulations, for the pulse propagating through the fast-light medium, superimposed with a companion pulse propagating through free-space over the same distance. (c,d) Zoomed-in view of: (c) the peak of the pulses and (d) their front discontinuities. These illustrative results show that, while the pulse peak can propagate superluminally, the front discontinuity is bounded by the speed of light in vacuum, thus obeying relativistic causality. As discussed in the text, the peak of a smooth pulse does not carry any genuine information that is not already present in the pulse front. As a result, since information cannot travel superluminally, an object cloaked by a three-dimensional fast-light medium, as proposed in Fig. 1. of Ref. \cite{2019Tsakmakidis}, would always be detectable using sufficiently accurate time-of-flight measurements.}
\end{figure*}

\begin{figure*}
 \centering
  \includegraphics[ width=0.65\linewidth, keepaspectratio]{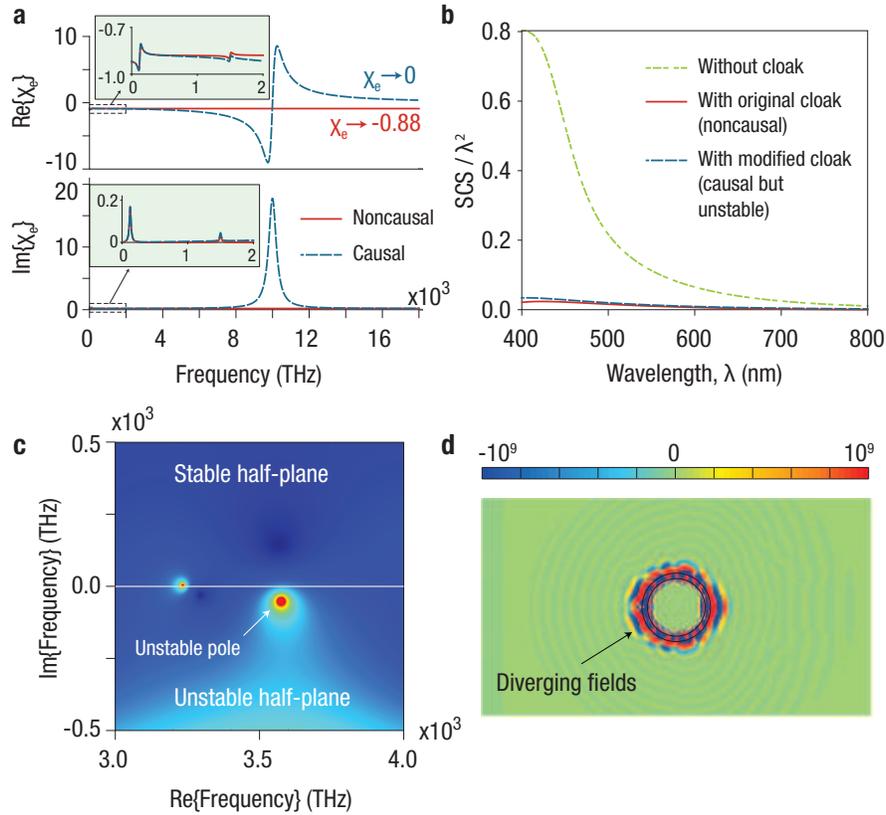}
\caption{\textbf{Physical limitations of invisibility cloaks made of fast-light media due to relativistic causality and stability}. 
(a) Electric susceptibility (top panel, real part; bottom panel, imaginary part) of the original fast-light (gain doublet) cloaking material in Fig. 2 of Ref. \cite{2019Tsakmakidis} (red solid curve) and a modified causal version of this material (blue dashed curve). The material models considered in Ref. \cite{2019Tsakmakidis} violate relativistic causality because the high-frequency limit of the susceptibilities is non-zero and negative. The main panels show the high-frequency limit and the insets show the frequency range of interest for cloaking operation. (b) Normalized scattering cross section (SCS) based on exact analytical calculations (Mie theory \cite{Bohren}) for the uncloaked and cloaked spherical dielectric objects considered in Fig. 2 of Ref. \cite{2019Tsakmakidis}. The plot shows the scattering spectra of both the original non-causal cloak, and the modified causal (but unstable) cloak, corresponding, respectively, to the red and blue curves in panel (a). (c) Magnitude of the scattering coefficient for electric dipole scattering from the causal cloaked sphere, plotted on the complex frequency plane, revealing the presence of an unstable pole. (d) Electric field distribution (time snapshot of the $x$-component) around the causal cloaked sphere on the $xz$-plane. The field profile is shown for a broadband pulse after the pulse has propagated for a duration of 4 fs, clearly showing the diverging fields around the cloaked object due to the onset of unstable (unbounded) oscillations. This implies that the low SCS for the cloaked cases in panel (b) is physically irrelevant since the cloaking material is either temporally noncausal or unstable. The scale bar is normalized to the maximum amplitude of the incident wave. Time-domain animation of the field distribution in (d) is given in Supplementary Video $1$$^{\dag}$.}
\end{figure*}



\cleardoublepage
\newpage
\onecolumngrid
\vspace{2cm}
\begin{center}
\large{  {\textbf{ Reply to ‘Reply to ‘Physical limitations on broadband invisibility\\ based on fast-light media’’}}}\\ \hfill
  
  \small{Mohamed Ismail Abdelrahman$^{1,2}$, Zeki Hayran$^{1,2}$, Aobo Chen$^{1}$, Francesco Monticone$^{1,*}$ \\ \hfill
  
\textit{$^{1}$ Cornell University, School of Electrical and Computer Engineering, Ithaca, New York 14853, USA}
  
\textit{  $^{2}$ These authors contributed equally.}
  
\textit{  $^{*}$ corresponding author: francesco.monticone@cornell.edu}} \\ \hfill
  
  \end{center}
  {{
This document provides additional results and comments supplementing the main criticisms raised in our
\textit{Matters Arising} [21] and addressing the main issues in the authors’ \textit{Reply} [22]. Specifically, we show that
the new cloak design provided in the \textit{Reply} [22] to our \textit{Matters Arising} is unstable, and we further point out
some issues in the authors’ claims. 

\begin{itemize}
\item We analyzed the new active cloak design, made of a causal fast-light material, that the authors of the original article provided in their \textit{Reply} [22] to our \textit{Matters Arising}, and we found that, similar to the
active cloak in Fig. 2, this new cloak is still unstable, as clearly evidenced by the presence of unstable
scattering poles and unbounded oscillations in its temporal response (see Fig. 3 and Supplementary
Video 2$^{\dag}$). As in our previous analysis, we used two fully independent methods to verify the unstable
response of the new cloak: exact Mie theory [20] (Fig. 3(a)) and FDTD numerical simulations using a
commercial software [23] (Fig. 3(b) and Supplementary Video 2$^{\dag}$). The unstable nature of this new cloak
reaffirms the difficulties and challenges in designing a broadband active cloak that is both physical and
free of any unbounded oscillations. This debate also highlights the importance, especially when dealing
with complex active media, of independently validating one’s results and claims using different
independent methods. Indeed, we are still puzzled that the authors of [22] never performed, neither in
the original article nor in their Reply, a truly independent validation of their results, for example in the
form of time-domain numerical simulations as in Fig. 3(b). 
  
  \item A contradiction worth reemphasizing is that, in their \textit{Reply} [22], the authors state they only performed
“frequency-domain calculations” using the material parameters that, as we pointed out in [21], are noncausal, whereas in the original article [1] they explicitly claimed, multiple times, to have performed
“full-wave causal simulations” (e.g., in the abstract) and to have provided “time-domain simulations
snapshots”. Certainly, frequency-domain simulations of a non-casual medium cannot be referred to as
“full-wave causal simulations”! Moreover, we are again puzzled that the authors of [1] never
performed, neither in the original article [1] nor in their \textit{Reply} [22], actual time-domain numerical simulations
of their proposed active cloaks (which would have clearly shown that their cloaks were either non-causal
or non-stable). In any case, causality and stability are key properties of a physical active system that
must be properly assessed and discussed regardless of the type of simulations employed in one’s study. 

\item Another relevant point to mention is that a “fast-light cloak” presented as in Fig. 1 of Ref. [1], where the
“extra pathlength is balanced out by the correspondingly larger group velocity of the pulse in the cloak”
would, in fact, require a transformation-optics based cloak [24], rather than a scattering-cancellation
based cloak as employed in Ref. [1]. Such a broadband transformation-optics cloak would require active
materials that are strongly anisotropic [24], thus making cloaks that work truly on the principle of
superluminal-group-velocity propagation around an object even more challenging to realize in practice.
\end{itemize}} 

\vspace{1cm}

\small{\noindent{[21] M. I. Abdelrahman, Z. Hayran, A. Chen, and F. Monticone. Physical limitations on broadband invisibility based
on fast-light media. \textit{Nat. Commun.} 12, 1-5 (2021). }

\noindent{[22] K. L. Tsakmakidis, O. Reshef, E. Almpanis, G. P. Zouros, E. Mohammadi, D. Saadat, F. Sohrabi, N. FahimiKashani, D. Etezadi, R. W. Boyd, and H. Altug. Reply to ‘Physical limitations on broadband invisibility based on fast-light media’. \textit{Nat. Commun.} 12, 1-3 (2021).} 

\noindent{[23] Lumerical FDTD Solutions, http://www.lumerical.com/. }

\noindent{[24] R. Fleury, F. Monticone, and A. Alù. Invisibility and Cloaking: Origins, Present, and Future Perspectives. \textit{Phys.
Rev. Appl.} 4, 037001 (2015).} }

\begin{figure*}
\centering
  \includegraphics[ width=0.9\linewidth, keepaspectratio]{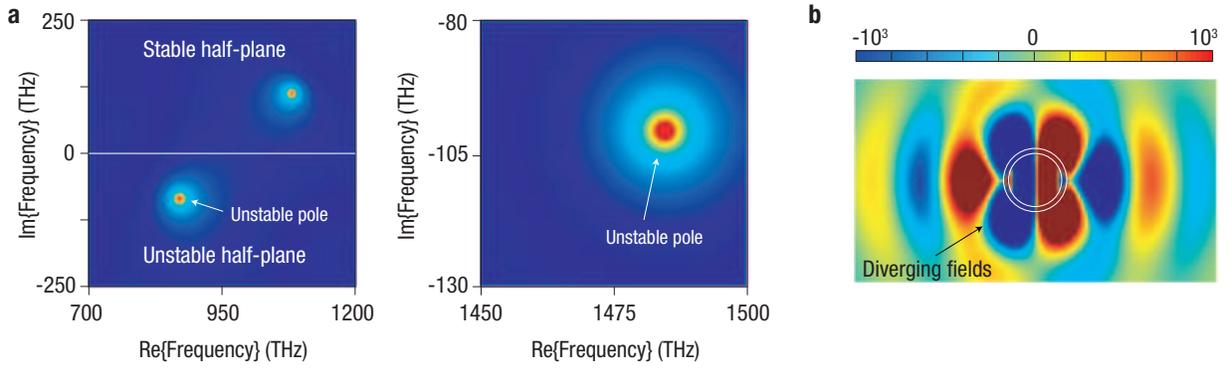}
\caption{\textbf{Unstable response of the active cloak provided in Ref. [22].}  (a) Magnitude of the scattering coefficient for
electric dipole scattering, in the complex frequency plane, for the cloak given in the \textit{Reply} to this \textit{Matters Arising} [22].
The left and right panels reveal the presence of two unstable poles in different frequency ranges. (b) Electric field
distribution (time snapshot of the x-component) around the cloaked sphere on the xz-plane, calculated with FDTD
numerical simulations using a commercial software [23]. The field profile is shown for the same broadband pulse as
in Fig. 2 after the pulse has propagated for a duration of 20 fs. The diverging fields around the cloaked object clearly
reveal the onset of unstable (unbounded) oscillations. The scale bar is normalized to the maximum amplitude of the
incident wave. Time-domain animation of the field distribution in (b) is given in Supplementary Video 2$^{\dag}$.}
\end{figure*}

\newpage

\noindent{$^{\dag}$ Online link to} \\
Supplementary Video 1: \url{http://bit.do/broadband-invisibility-limits-supp-video1} \\
Supplementary Video 2: \url{http://bit.do/broadband-invisibility-limits-supp-video2}  \\

\end{document}